\documentclass[12pt, onecolumn, draftcls,]{IEEEtran}
 
 \usepackage[cmex10]{amsmath}
\usepackage{amssymb}
\usepackage{amsthm}
 
\usepackage[latin1]{inputenc} 
\usepackage{tikz}
\definecolor{gold}{rgb}{0.85,.66,0}

\begin{document}

\title{NOMA Systems Optimization to Ensure Maximum Fairness to Users}

\author{Jaime L. Jacob and Taufik Abrao\\
\normalsize Department of  Electrical Engineering (DEEL)
State University of Londrina  (UEL). \\
Po.Box 10.011, CEP:86057-970, Londrina, PR, Brazil. \\

\thanks{J. L. Jacob and T. Abrao are with the Department of  Electrical Engineering (DEEL)
State University of Londrina  (UEL). Po.Box 10.011, CEP:86057-970, Londrina, PR, Brazil.  Email: taufik@uel.br}
}
\maketitle

\begin{abstract}
In this contribution, the optimization of power proportion allocated for each user in the downlink (DL) non-orthogonal multiple access (NOMA) systems have been developed. Successive interference cancellation (SIC) technique recovers users' signal with high difference between channel gains, in order to find the lowest optimum power proportion {required to guarantee} an equal data rate for all active users. Following the same approach, the optimum power proportion for each user and also the minimum total power to achieve the same rate for all users (maximum fairness) were obtained as a design goal. Moreover, the same design methodology was developed seeking to maximize the NOMA system energy efficiency (EE). It was possible to find the {maximum EE point} and the respective power distribution among the users for a certain circuitry power consumption. For all NOMA users in which the optimal operation point for EE maximization was parameterized, it was possible to find the total power, power ratio and the equal rate values. As a result, one can find the tradeoff point between EE and sum rate for each system scenario, as well as the best resource efficiency operation point. By considering the same rate for all users, the system attains maximum fairness among the users.
\end{abstract} 

\begin{IEEEkeywords}
Non-orthogonal multiple access (NOMA), Optimization, successive cancellation of interference (SIC), energy efficiency (EE), fairness.
\end{IEEEkeywords}


\maketitle


\section{Introduction}
Non-orthogonal multiple access (NOMA) system is a promising technique for 5G technology by optimizing bandwidth usage where users can share it, since they have different power ratings. As a result, users at the edge of the cell will be treated under the same conditions as cell users themselves in wireless cellular systems \cite{Tse_Viswanath_2005}. To reach this goal, some studies {have been done analyzing the necessary} conditions for the use of NOMA system. {For instance} one can maximize the {symbol transmission rate} of two users in a cluster \cite{Kim2013} {aiming at maximizing} the sum capacity of them {by deploying} beamforming technique in the base station (BS). However, is two users per cluster the ideal configuration? Additionally, the feedback rank is used to determine the NOMA application in cases where users with the best rank are selected while other users continue {using} the orthogonal multiple access (OMA) \cite{Chen2014}. Note that those authors do not consider the optimal power distribution. Instead of {setting a single} BS, {one} can use two BS and apply the Alamouti code combined with the successive interference cancellation (SIC) technique to recover the signal at reception \cite{Choi2014}. 

Another approach {to improve NOMA capacity consists in} separate the users into three categories (singletons, weak users, and strong users) and to analyze the transmission rate and SINR for the three categories, emphasizing the Quality of Service (QoS) of weak users \cite{Dhakal2019}. {such} work analyzed only two users per cluster. Furthermore, other authors \cite{Zhang2017} developed EE optimization and got the green point, i.e., the optimum system operation point that maximizes the overall EE, but they did not consider fairness among users. The tradeoff between EE and spectral efficiency (SE) was studied {in} \cite{Liu2017}, but the authors did not analyze different numbers of users. Moreover, the maximization of the minimum rate for each user was analyzed in \cite{Bui2019}, but despite mentioning fairness such concept was not systematically explored. {Besides,} the trade-off between EE and SE {considering} fairness {has been analyzed in \cite{Dadong_2018}}, but it did not guarantee maximum fairness {among the} users, and the calculation used the traditional concept of Jain's fairness \cite{Jain1984}. {Currently, the} concept of fairness based on the considerations of information theory \cite{Gui2019} {has become popular}.

A research aimed to compare the uplink {(UL)} NOMA system with OMA using resource allocation fairness as a reference and {considering} only Jain's index {was reported in \cite{Wei2017}}. The authors analyzed fairness for only two users. In another research \cite{PengXu2017} an optimal power allocation scheme in the system was investigated from the concept of $\alpha$-fairness with the solution of the ergotic rate maximization and the sum of capacity problem in a downlink {(DL)} NOMA system. A waterfilling-based power allocation technique has been applied to the DL NOMA combined with proportional fairness scheduler to maximize the average rate with near-optimal power distribution between subbands and high fairness \cite{Marie2017}. In other analysis, the trade-off between EE and SE from the combination of OFDMA and NOMA systems were made from the user's fairness restriction \cite{Yan2018}. Resource efficiency for different weights between EE and SE and different maximum transmission powers were analyzed. Another paper \cite{Xing2018} analyzed the average sum rates, together or not, with the maximization of the sum of the delay-limited rates. This analysis is subject to user fairness in order to obtain optimal power control. This is all done for a two-user {DL} NOMA system over a fading channel. {Research in \cite{Qi2018} has} addressed {the} power allocation problem that maximizes the minimum rate obtained among users. Additionally, by simulation, the minimum user rate was compared {in terms of} the Jain's fairness index for different power allocation schemes in NOMA {versus} OMA systems. {Authors in \cite{Muhammed2019} have analyzed the jointly} optimization of energy efficiency and fairness among users with respect to subcarriers and the power allocation of a {DL} multi-carrier (MC) NOMA system. The authors compare the EE for MC-NOMA with equal power, difference of convex programming \cite{Fang2016} and their proposed convex programming sequence for the best link and the worst link and combination. {Furthermore, a} power allocation {strategy for} two users in a {UL} NOMA system {with} proportional fairness {guarantees has been proposed in \cite{LiangChen2019} where} two scenarios analyzed: basic scenario users are distributed {inside} the cell, and in the complex scenario {where} the interfering users are {placed} outside the cell.

\noindent{\it Contribution}: This work establishes the best power allocation policy in NOMA system in terms of sum-power allocation coefficients which guarantees {of equal} rate (fairness approach) for all users. For that, we deploy convex optimization techniques \cite{Boyd2004}, \cite{Bertsekas2009} to solve the optimization problem.  Differently of \cite{Timotheou2015}, in which the power proportion {allocated} for each user is found, herein we have further considered the energy efficiency maximization and the fairness among users in terms of equal data rate guarantee. Indeed, this paper also considers the EE-NOMA optimization problem by finding the minimum total power needed to reach a specified equal rate per user. From the jointly EE and sum rate optimization with maximum fairness guarantee {optimization problem}, we have find the best EE-SE trade-off and thereby {obtain} get the best {NOMA} system {configuration in terms of} resource efficiency.

\section{NOMA System}\label{Model}
\begin{figure}[!htb]
	\centering\includegraphics[width=0.75\textwidth]{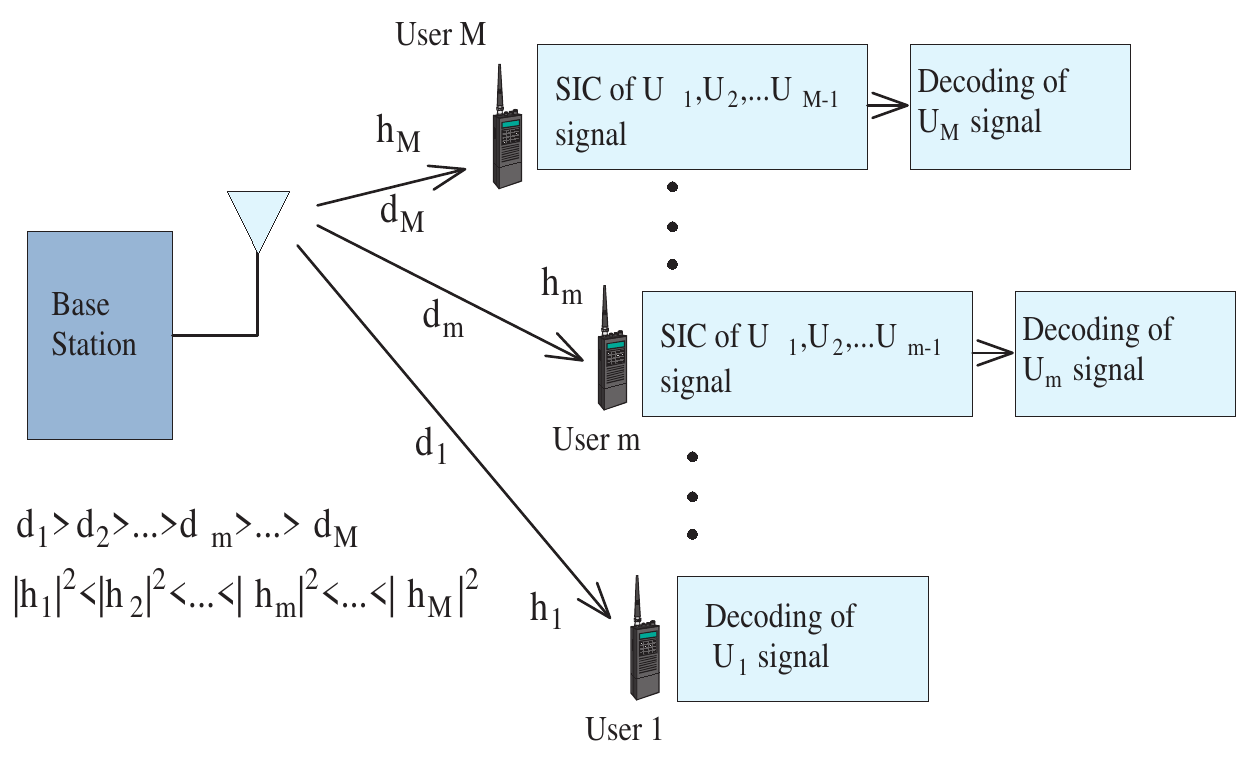}
	\caption{NOMA with $M$ users.}\label{fig:NOMAcomMusuarios}
\end{figure}

Consider a cellular system with down-link transmission scenario, in which one base station (BS) located in the center of the disc with radius $\Re_D$ and $M$ users aligned, Figure \ref{fig:NOMAcomMusuarios} . All terminals are equipped with a single antenna.
The channel between the users and the base station \cite{Ding2014} is given by \footnote{It is adding 1 in the channel gain denominator, eq. \eqref{Channel}, it is warranted that at the $0$ distance, the transmitted signal is equal to the received signal. Indeed, this is one of  possible pathloss models found in the literature for instance, in \cite{Inaltekin2009}.}
\begin{equation}
h_i=\frac{g_i}{\sqrt{1+d_i^{\alpha}}}\;\;\;\rm for\;\;i=1,2,\cdots M
\label{Channel}
\end{equation}
where $g_i$ represents the channel gain with Rayleigh fading, $\alpha$ is the path loss factor, and $d_i$ represents the distance from de user to the base station {\cite{Inaltekin2009}}. Without loss of generality, the channels are sorted as $|h_1|^2\leq|h_2|^2\leq\cdots\leq|h_M|^2$. The transmission signal in the BS with application of the NOMA technique is given by
\begin{equation}
x=\sum_{i=1}^M\sqrt{\beta_iP}s_i
\end{equation}
where $s_i$ is message from {$i$-th} user, $P$ is the transmission power, and $\beta_i$ is the power allocation coefficient, i.e. $\beta_1\geq\beta_2\geq\cdots\geq\beta_M$. There are the superposition of $M$ signals to its $M$ users via power-domain division. The signal
received at the {$m$-th} subscriber is given by
\begin{equation}\label{GenericReceived}
y_m = h_m\sum_{i=1}^M\sqrt{\beta_iP}s_i + z_m
\end{equation}
where $z_i$ is i.i.d. additive white complex Gaussian noise (AWGN) with zero mean and {variance} $\sigma_n^2$. In the system model presented here, only user $1$ does not use the SIC technique in signal detection. The SIC technique is initially used in user $2$, where user $1$ has its interference fully extracted and the other users are treated as noise. The privilege of the user closest to BS, that is, the {$M$-th} user is to have the interfering signals fully extracted. To calculate the capacity, it is first necessary to calculate the received signal-to-interference-plus-noise power ratio (SINR). From equation (\ref{GenericReceived}) we can obtain the generic SINR for the $1\leq m\leq(M-1)$ given by 
\begin{equation}\label{eq:gamma}
\gamma_m=\frac{P|h_m|^2\beta_m}{P|h_m|^2\sum_{i=m+1}^M\beta_i+\sigma_n^2}, \qquad m=1\ldots, M-1
\end{equation}
and for the last (near-BS) user results:
\begin{equation}
\gamma_M=\frac{P|h_M|^2\beta_M}{\sigma_n^2}
\end{equation}

The channel capacity is given by $\log_2(1 + \rm SINR)$, so the capacity of users are given by \cite{Ding2014}
\begin{equation}\label{nearUSERS}
R_m=\log_2\left(1+\gamma_m\right)\qquad m=1,\ldots M \qquad [\rm bits/s/Hz].
\end{equation}

Herein, it is assumed that the interference from the first user has been completely canceled when using the SIC technique.

Energy efficiency (EE) is given by the ratio between the sum of rates achievable by all users in system and the total consumed power, given by
\begin{equation}\label{EE}
{\rm EE}=\frac{R_s}{P_t+MP_c} = \frac{MR}{P_t+MP_c},\qquad \left[\frac{\rm bits}{\rm J \cdot Hz}\right]
\end{equation}
where $R_s=\sum_{m=1}^M R_m=MR$ is the sum of the rates transmitted by all the users given by  Eq. \eqref{nearUSERS} where all uses have the same capacity, {\textit{i.e.}}, $R_m=R$; the total RF power consumed is given by:
\begin{equation}\label{eq:Pt}
P_t=\varrho\cdot \sum_{i=1}^M\beta_iP \qquad [\rm W] 
\end{equation}
where $\varrho\geq1$ is the RF power amplifier inefficiency, assumed equal for all of them feeding each Tx antenna;  $\beta_i$ is the power allocation coefficient, and $P_c$ is the circuitry power required to signal processing at the transmitter side, assumed be fixed value per user.

\section{{Downlink NOMA Optimization}}\label{Optimization}

In this section, the optimization of the NOMA system is done in two ways. In the same optimization process, the goal is to get the minimum proportion of power for each user {needed to attain a specific} target equal rate (maximum fairness), and still lower total transmit power to obtain {such} same minimum  {fair data rate. This goal is achieved by simultaneously maximizing the EE via power transmitting optimization and the} proportion of power among users. 

{In the sequel, we discuss the present the fairness index, discuss the solution for the ratio of power optimization problem, as well as the power allocation solution under the perspective of EE optimization. Finally, the concept of resource efficiency is revisited.} 

\subsection{Fairness {Indexes}}\label{sec:fairness}
Fairness index can be obtained by the formula based on the {\it Jain's fairness index}  \cite{Jain1984, Wang2017, Wang2017a}{, as well as making considerations based on information theory. Hence, from the conventional definition of fairness,  the Jain's fairness index can be written as}

\begin{equation}\label{fairness}
\mathcal{F}_\textsc{j}=\frac{\left(\sum\limits_{k=1}^MR_k\right)^2}{M\sum\limits_{k=1}^MR_k^2}.
\end{equation}
In the optimization design context of this paper the same rate has been adopted for all the users; hence, it can be verified by eq. \eqref{fairness} that {under this condition, the} calculated fairness results in $\mathcal{F}_\textsc{j}=1$ being its maximum value. For illustration purpose, let us consider the following toy example: assuming $M=4$ users, with $R_1=R_2=R_3=1$ and $R_4=0.7$,  resulting in $(\sum_{k=1}^4R_k)^2=13.69$ and $4\cdot\sum_{k=1}^4R_k^2=13.96$; so, $\mathcal{F}_\textsc{j}=0.98$ results very close to $1$, since only one user presents data rate different than one.

One alternative way to define fairness is based purely on {\it information theory} considerations, as discussed in \cite{Gui2019}, being expressed by:
\begin{equation}\label{NewFairness}
\mathcal{F}_\textsc{it}=1-\frac{\hat{\Delta}^2}{\hat{R}^2}
\end{equation}
were $\hat{\Delta}^2=\frac{1}{M-1}\sum\limits_{m=1}^M\Delta_m^2$ is the unbiased estimate of the mean-square rate deviations and $\hat{R}^2=\frac{R_s^2}{M}$ is average square sum rate. The $\Delta_m^2$ is the squared deviations of the actual user rates from their fair rate {($R^{\rm f})$} as follows:
\begin{equation}
\Delta_m^2=\left(R_m^{\rm f}-R_m\right)^2,\qquad m=1,2,\dots,M
\end{equation}
The fair rate for {the $m$th user is defined as} 
can be represented by
\begin{equation}
R_m^{\rm f}=\log_2\left( 1+\frac{P\left|h_m\right|\beta_m}{\sigma_n^2}\right)\cdot\frac{R_s}{R_c}
\end{equation}
where the cumulative rate is given by
\begin{equation}
R_c=\sum\limits_{m=1}^M\log_2\left( 1+\frac{P\left|h_m\right|\beta_m}{\sigma_n^2}\right).
\end{equation}
The authors {in} \cite{Gui2019} justify the cumulative rate as the sum rates of an OMA system that allows a separate channel for each user. Calculating fairness in this way, {the index of eq. \eqref{NewFairness}} reaches the maximum value of $1$ {when both condition are achieved, a uniform power distribution and the equal rate for all users. For instance,} when a user makes full use of the network resources, {the $\mathcal{F}_\textsc{it}$} index reaches its minimum value of $0$.

\subsection{Power Optimization {Problem}}\label{PowerOptimization}

The  {relative power ratio $\beta_i$ as well as the minimum total power $P$ to} be considered for each user to attain a minimum equal data rate for all users (maximum fairness) {can be obtained solving the follows optimization problem}:
\begin{eqnarray}\label{ConstraintsOptimizationBETA}
\begin{array}{rl}
\min\limits_{(P,\beta_i) \in \Re^+} & \quad \sum_{i=1}^M\beta_iP\\
\text{s.t.}\;\;\; {\text{(C.1)}}       & \quad\sum_{j=1}^M\beta_j=1  \\
		{\text{(C.2)}}       & \quad \beta_j>0; P>0; R>0  \\
		{\text{(C.3)}}       & \quad \beta_m=\left({2^R-1}\right)\sum\limits_{i=m+1}^M{\beta_i +\left({2^R-1}\right)\frac{{\sigma_n^2}}{{P\left|{h_m}\right|^2}}}\;\;\text{to}\;\;m=1,\cdots,M-1\\
{\text{(C.4)}}       & \quad \beta_M =\left({2^R-1}\right)\frac{{\sigma_n^2}}{{P\left|{h_M }\right|^2}}  \end{array}
\end{eqnarray}
where the expression for the proportions of total power {can be derived from \eqref{eq:gamma}--(\ref{nearUSERS})}.

One need to prove that the set of constraints {in \eqref{ConstraintsOptimizationBETA} is} convex. For this, the restriction {(C.1)} can be developed by replacing the restrictions {(C.3)} and {(C.4)} in the constraint {(C.1)} subjected to the constraints of {(C.2); hence, the following expression} is obtained.
\begin{equation}\label{restricoes} 
f=\sum\limits_{m=1}^{M-1}{\left({\left({2^R-1}\right)\sum\limits_{i=m+1}^M{\beta_i+\left({2^R-1}\right)\frac{{\sigma_n^2}}{{P\left|{h_m}\right|^2}}}}\right)}+\left({2^R-1}\right)\frac{{\sigma_n^2}}{{P\left|{h_M }\right|^2}}-1
\end{equation}
The Hessian of (\ref{restricoes}) must be semi-definite positive, that is, ${\nabla}^2f\succcurlyeq0$ to be convex \cite{Boyd2004}. Therefore the eigenvalues of the Hessian must be greater than or equal to zero. After applying the {Hessian and eigenvalues definition in (\ref{restricoes}), one can obtain} the following polynomial characteristic {for the eigenvalues determination}:
\begin{equation}
-\lambda^M+\frac{2\left({2^R-1}\right)\sigma_n^2\lambda^{M-1}}{\sum\limits_{m=1}^{M-1}\left|{h_m}\right|^2P^3}+\frac{2\left({2^R-1}\right)\sigma_n^2\lambda^{M-1}}{P^3\left|{h_M }\right|^2}=0
\end{equation}
{Hence, taking} $\lambda^{M-1}$ in evidence, so that there are $M-1$ eigenvalues equal to zero, and therefore {one can obtain:}
\begin{equation}
\lambda=\frac{2\left({2^R-1}\right)\sigma_n^2}{P^3}\left(\frac{1}{\sum\limits_{m=1}^{M-1}\left|{h_m}\right|^2}+\frac{1}{\left|{h_M }\right|^2}\right)
\end{equation}
Since the variable $P$ is positive, as well as the constants $R$, $\sigma_n^2$ and $\left|{h_i}\right|^2$, the eigenvalues {associated to \eqref{restricoes}} are positive or equal to zero and the equation in question is semi-definite positive, then the constraints {(C.1)--(C.4) constitute}  a convex set.

\begin{figure}[!htbp]
	\centering
	\includegraphics[width=0.48\textwidth]{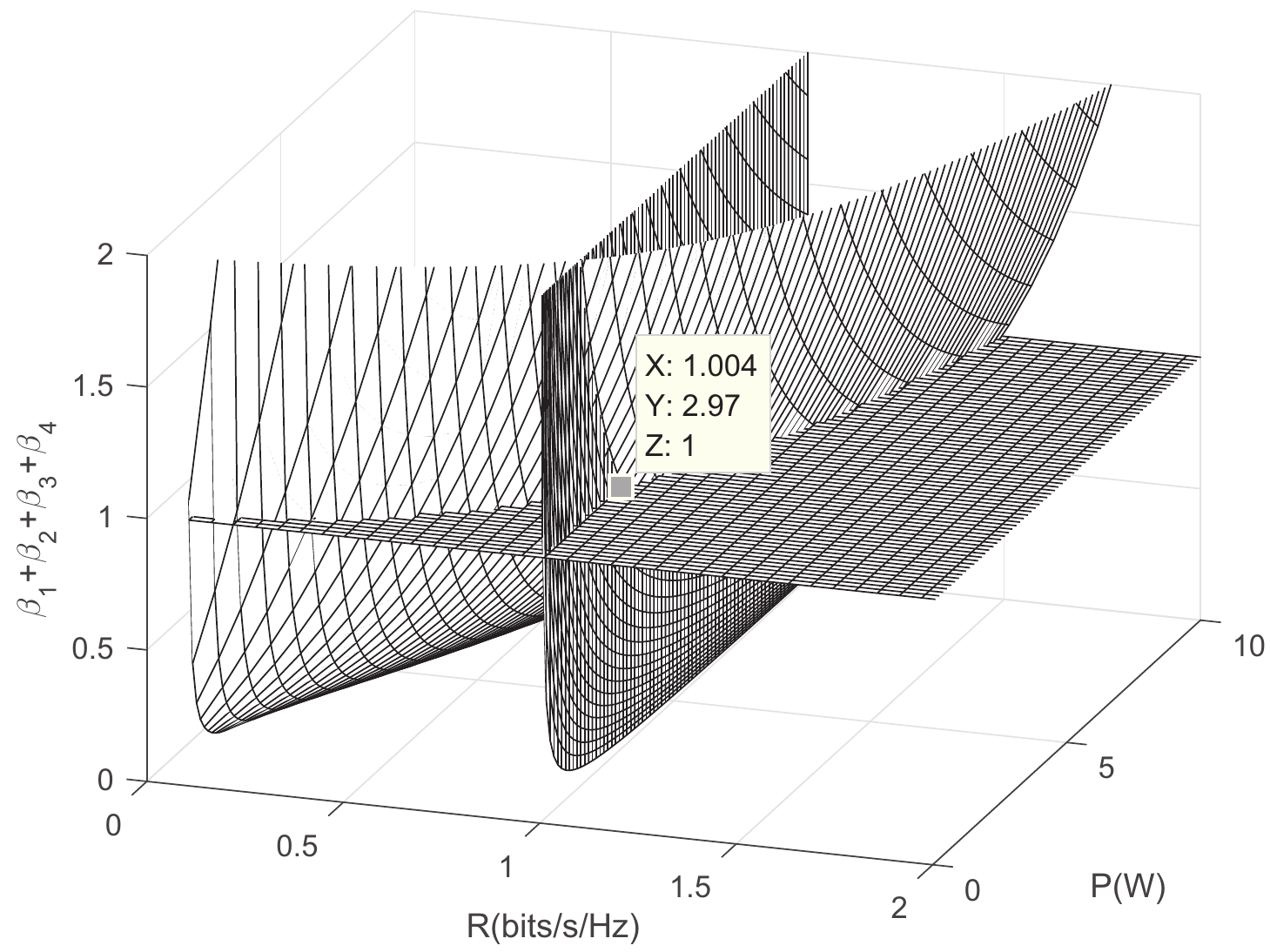}
	\hspace{2mm}
	\includegraphics[width=0.49\textwidth]{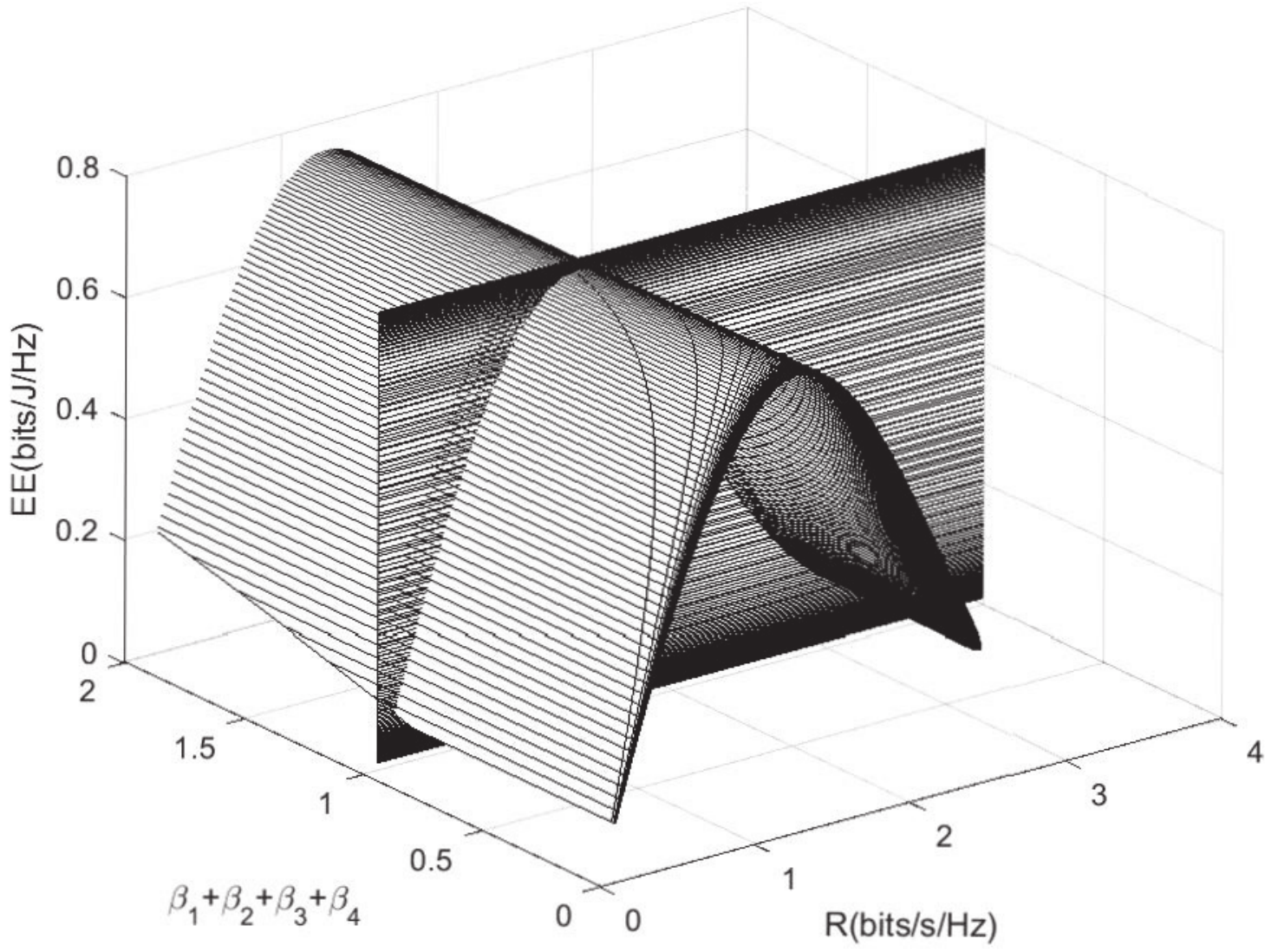}\\
	\footnotesize (a) $\beta$'s $\times$ Rate $\times$ Power (c3, c4 of eq. \eqref{ConstraintsOptimizationBETA}). \hspace{10mm} (b) EE $\times$ $\beta$'s $\times$ Rate, with $P_c=1$W eq. \eqref{EE} and (c3, c4 of eq. \eqref{ConstraintsOptimizationBETApReEE}).
	\vspace{-2mm} 
	\caption{Closed set for $\sigma_n^2=1\mu$W and $M=4$ users with distances $d_1=1.0$km, $d_2=0.7$km, $d_3=0.5$km and $d_4=0.3$km.}
	\label{fig:ConvexSet4BETASxPxR}
\end{figure}

The constraints form a collection of convex sets and the intersection of these sets is convex. According to Fig. \ref{fig:ConvexSet4BETASxPxR}.(a) the constraints defined in (\ref{ConstraintsOptimizationBETA}) form a closed set; in this way, the presented optimization problem has a solution with optimal global. In Fig. \ref{fig:ConvexSet4BETASxPxR}.(a) one can observe the subspace given by the condition of $\beta_1+\beta_2+\beta_3+\beta_4=1$ and by the rate of each user $R=1$ bits/s/Hz.  In this case, the solution is obtained for a total power $P\approx 2.97$ W. However, the power proportion values can not be obtained so easily {in real practical interest system and channel scenarios.} Moreover, Fig. \ref{fig:ConvexSet4BETASxPxR}.(b) depicts the set of solutions for the EE optimization problem. This set is formed by the sum of the power proportions equal to $1$, i.e., $\beta_1+\beta_2+\beta_3+\beta_4=1$ and the intersection of the other plane defined by the EE and user's rate. 

\subsection{Optimization of User Power Allocation from the EE Perspective}\label{PowerOptimizationEE}

The equation (\ref{EE}) is quasi-concave \cite{Zhang2017}. We use the Dinkelbach method to turn a quasi-concave objective function into a concave one \cite{JieTang2019}, as indicated in Table \ref{table:AlgoritmoDinkelbachPeR} to solve problem \eqref{ConstraintsOptimizationBETApReEE} , where $U_R=MR$ and $U_T=\varrho\sum_{j=1}^M\beta_jP+MP_c$.
\begin{center}
	\begin{table}[!htbp]
		\centering
		\caption{Iterative power allocation and data rate and/or power allocation algorithm based on Dinkelback method.}
		\begin{tabular}{l}
			\hline
			\textbf{Dinkelbach's algorithm}\\
			\hline
			1) Let $n=0$ and $q^{(n)}=0$; \\
			\hspace{4mm}Set $\varepsilon>0$ as the stopping criterion; \\
			2) \textbf{REPEAT} \\
			\begin{tabular}[c]{@{}l@{}}3)  \hspace{2mm}For a given $q^{(n)}$, solve \eqref{ConstraintsOptimizationBETApReEE} and/or \eqref{ConstraintsOptimizationBETApEE} to obtain first the power\\  
				\hspace{5mm}allocation and data rate $P^{(n)}$ and $R^{(n)}$ and/or second the power allocation $P^{(n)}$;\end{tabular} \\
			\hspace{4mm}\textbf{IF} $U_R(P^{(n)},R^{(n)})-q^{(n)}U_T(P^{(n)},R^{(n)})\leq\varepsilon$ and/or $U_R(P^{(n)})-q^{(n)}U_T(P^{(n)})\leq\varepsilon$ \\
			5)    \hspace{6mm}Convergence = \textbf{TRUE}; \\
			6)    \hspace{6mm}\textbf{RETURN}
				$P^*=P^{(n)},R^*=R^{(n)},q^*=q^{(n)}$ and/or $P^*=P^{(n)},q^*=q^{(n)}$; \\
			7)   \hspace{4mm}\textbf{ELSE}\\
			8)    \hspace{6mm}Convergence = \textbf{FALSE}; \\
			9)    \hspace{6mm}Set $n=n+1$ and $q^{(n)}=\frac{U_R(P^{(n-1)},R^{(n-1)})}{U_T(P^{(n-1)},R^{(n-1)})}$ or $q^{(n)}=\frac{U_R(P^{(n-1)})}{U_T(P^{(n-1)})}$; \\
			10)   \hspace{3mm}\textbf{END IF} \\
			11) \textbf{UNTIL} Convergence = \textbf{TRUE}.\\
\hline
		\end{tabular}
		\label{table:AlgoritmoDinkelbachPeR}
	\end{table}
\end{center}

 The EE optimization problem to obtain the optimal power allocation $P$, data rate $R$ and power proportion $\beta_i$ {for all the users can be formulated as:}
\begin{eqnarray}
\label{ConstraintsOptimizationBETApReEE}
\begin{array}{rl}
\min\limits_{(P,\beta_j,R) \in \Re^+} &\quad -\left( MR-q^n\left(\varrho\sum_{j=1}^M\beta_jP+MP_c\right)\right) \\
\text{s.t.}\;\;\;{\text{(C.1)}}   & \quad \sum_{j=1}^M\beta_j=1  \\
{\text{(C.2)}}   & \quad      \beta_j>0; P>0; R>0  \\
		{\text{(C.3)}}   & \quad  \beta_m=\left({2^R-1}\right)\sum\limits_{i=m+1}^M{\beta_i+\left({2^R-1}\right)\frac{{\sigma_n^2}}{{P\left|{h_m}\right|^2}}}, { \quad \text{for} \;\;m=1,\cdots,M-1 }\\
             {\text{(C.1)}}   & \quad  \beta_M=\left({2^R-1}\right)\frac{{\sigma_n^2}}{{P\left|{h_M}\right|^2}} \end{array}
\end{eqnarray}

By analyzing the convexity of the constraints {regarding the total power $P$, proportion of the total power among users $\beta_i$ and data rate $R$, one can conclude that the set (C.1)--(C.4) in \eqref{ConstraintsOptimizationBETApReEE} are} non-convex, since the eigenvalues of the Hessian with these variables result complex values. 

However, by setting the data rate $R$ while maintaining the total power $P$ and the proportion of the total power $\beta_i$ among users as {optimization variables,  the problem in \eqref{ConstraintsOptimizationBETApReEE} becomes convex since } \eqref{restricoes} is obtained. As proved earlier this set of constraints with {such variables is convex. Hence the problem \eqref{ConstraintsOptimizationBETApReEE} is reformulated as  \eqref{ConstraintsOptimizationBETApEE}} . We use the Dinkelbach method to turn a quasi-convex objective function into a convex one, as Table \ref{table:AlgoritmoDinkelbachPeR} to the problem \eqref{ConstraintsOptimizationBETApEE}.

\begin{eqnarray}
\label{ConstraintsOptimizationBETApEE}
\begin{array}{rl}
\min\limits_{(P,\beta_j) \in \Re^+} & \quad  MR-q^n\left(\varrho\sum_{j=1}^M\beta_jP+MP_c\right) \\
\text{s.t.}\;\;\;\ & \quad {\text{(C.1), \,\, (C.2), \,\, (C.3), \,\, (C.4)}}             \end{array}
\end{eqnarray}

\subsection{Resource Efficiency}\label{sec:ResourceEfficiency}
In conventional system designs instead of focusing on the SE or the EE separately, it is much more effective balancing the attainable system SE and EE by adopting resource efficiency (RE) metric \cite{Zhang2017}. The SE is defined for a single cell system with $M$ single-antenna users as $\text{SE}=\sum_{i=1}^MR_i$. The resource efficiency is expressed as a weighted sum of the EE and the SE can be formulated as:
\begin{equation}\label{RE}
\text{RE}=\xi_0\text{EE}+\text{SE}
\end{equation}
where $\xi_0$ is the weighting factor in [W] controlling the weights of EE and the SE on the design. Hence, when $\xi_0=0$ the expression \eqref{RE} reduces to the SE.

\section{Numerical Results}

The parameter values used in the simulations are described in Table \ref{table:ParametersSimulation} . The tool used for the optimization was the MatLab\textsuperscript{\textregistered} \textit{fmincon} function with the technique sequential quadratic programming (SQP).
\begin{center}
	\begin{table}[!htb]
		\centering
		\caption{Adopted Parameters for the Simulation Scenarios.}
		\begin{tabular}{ll}
			\hline
			\textbf{Parameter} & \textbf{Value} \\
			\hline
			Users number          & $M=2$ to $12$ \\
			Path-loss exponent    & $\alpha=[2.0;\;\, 3.0;\,\, 4.5]$ \\
			Power of white noise  & $\sigma_n^2=0.1\mu\rm W$ \\
			Power Budget          & $P=120$ W\\
			Target equal use's rate & $R\in [0.1;0.5;1;1.5;2;2.5;3]$ [bit/s/Hz]\\
			Circuits Power        & $P_c=250 \,\, \left[\frac{\rm mW}{\rm user}\right]$ \\
			RF power amplifier inefficiency & $\varrho=1.4$ \\
			Number of samples     & $10^4$ \\
			Monte-Carlo  simulation Trials & $5$\\
			\hline
			\multicolumn{2}{c}{{\bf Scenarios}: User distribution inside the cell}\\
			\hline
			Scen. $1$ -- $\sum d_i=1.0$ \,\,\, [km] & $d_1=0.34,\;d_2=0.29,\;d_3=0.22,\;d_4=0.15$ \\
			Scen. $2$ -- $\sum d_i=2.0$  \,\,\, [km] & $d_1=0.8,\;d_2=0.6,\;d_3=0.4,\;d_4=0.2$ \\
			Scen. $3$ -- $\sum d_i=4.0$ \,\,\, [km]   & $d_1=1.8,\;d_2=1.2,\;d_3=0.7,\;d_4=0.3$ \\
			Scen. $4$ -- $\Re_D$ [m] (uniform. distrib.) & $50,\;100,\;200,\;300,\;400$ \\
			\hline
			\multicolumn{2}{c}{\bf Power Allocation Strategies}\\
			\hline
			Inverse power allocation (ICA) & $P_i$ defined by eq. \eqref{channelInverse}\\
			Equal-rate optimal power allocation  (ERPA) & subsection \ref{sec:fairness} \\
			\hline
		\end{tabular}
		\label{table:ParametersSimulation}
	\end{table}
\end{center}
 The SQP algorithm \cite{Nocedal2006} is a powerful iterative method for nonlinear optimization and non-convex systems \cite{Ding2015b}. This algorithm can be considered as one of the best nonlinear programming method for constrained optimization problems. It outperforms every other nonlinear programming method in terms of efficiency, accuracy and percentage of successful solutions over a large number of test problems \cite{Elaiw2012}. The SQP algorithm converges to the local minimum \cite{Elaiw2012} but very close to the global minimum. It was considered a range of values from the lowest to the highest to obtain the behavior of the system with different number of users, path loss exponent values, target equal user's rate and $\Re_D$. In this way, one can trace a system's behavior considering different practical scenarios of interest. Unlike Scenario $4$, where users have a uniform distribution for the respective $\Re_D$, in the Scenario $1$, $2$ and $3$ users have a fixed distance from BS.
\begin{figure}[!htb]
	\centering
	\includegraphics[width=0.650\textwidth]{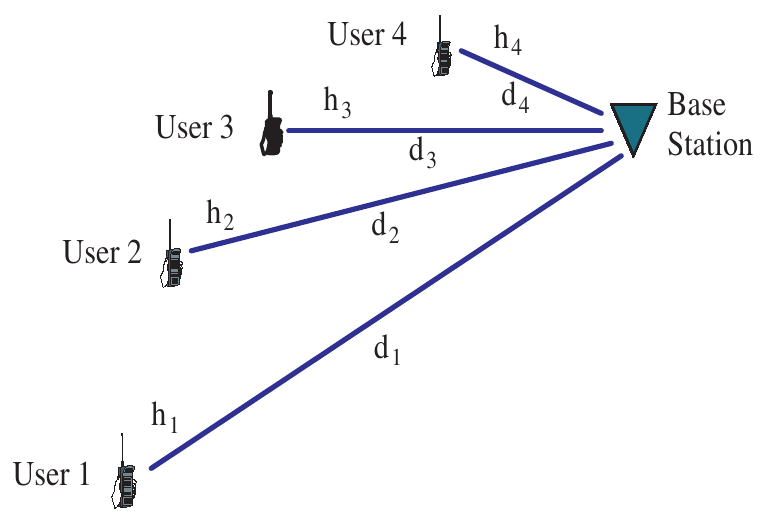}
	\vspace{-3mm}
	\caption{Distribution of users in a NOMA system with $4$ users.}
	\label{fig:CelulaNOMA}
\end{figure}

Fig. \ref{fig:CelulaNOMA} shows the distribution of users relative to the base station in a cell. We used the power assignment technique among users proportional to the inverse of the channel response (ICA) \cite{El-Sayed2016} to compare the sum rates with egalitarian rates among all users, given by:
\begin{equation}\label{channelInverse}
P_i=\frac{\sum_{k=1}^M\left|h_k\right|^2}{\left|h_i\right|^2}P,\,\, \qquad i=1,\ldots, M.
\end{equation}

\subsection{ERPA $\times$ ICA Performance}
\begin{figure}[!htb]
	\centering
	\includegraphics[width=0.487\textwidth]{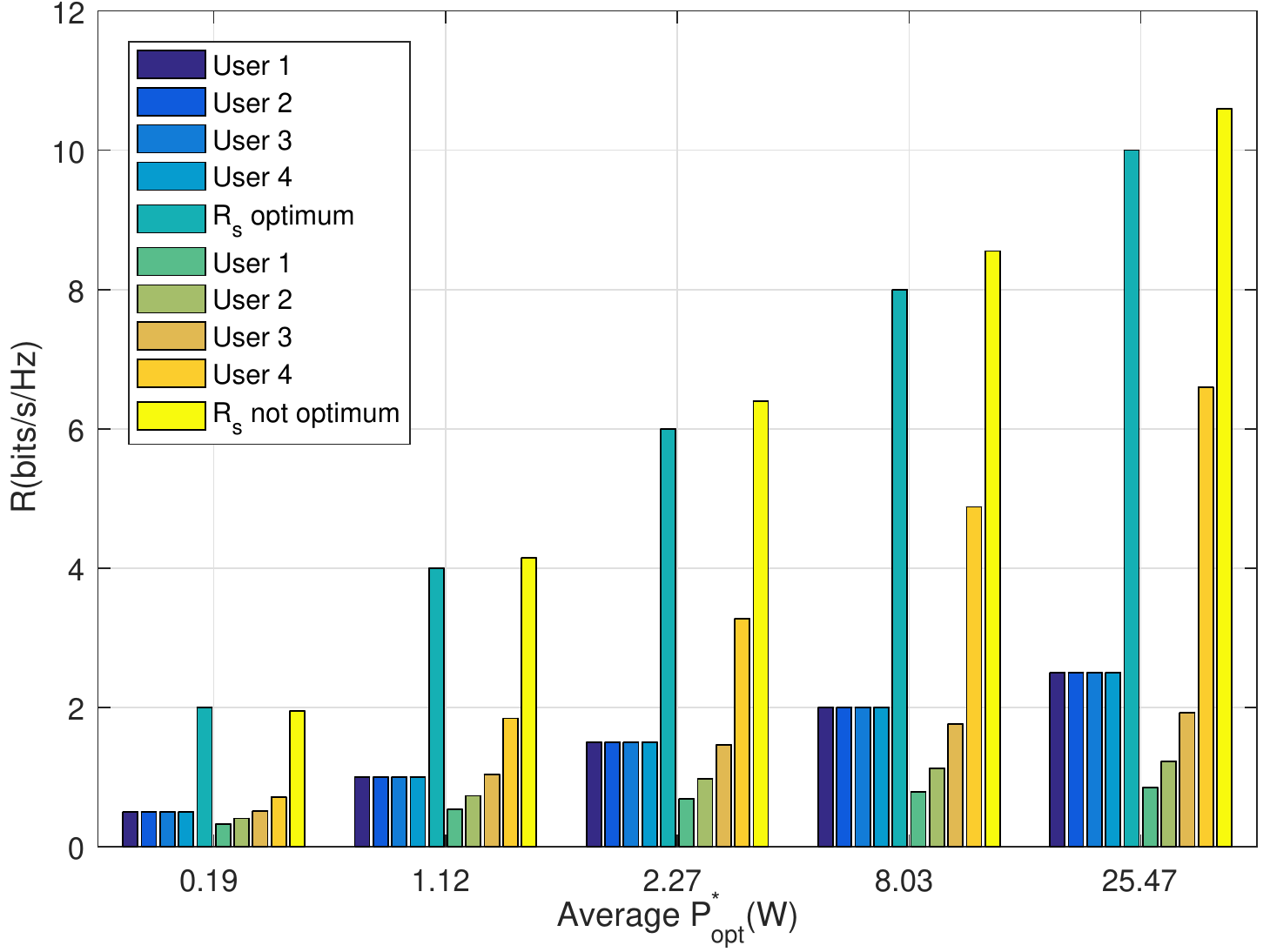}
	\includegraphics[width=0.487\textwidth]{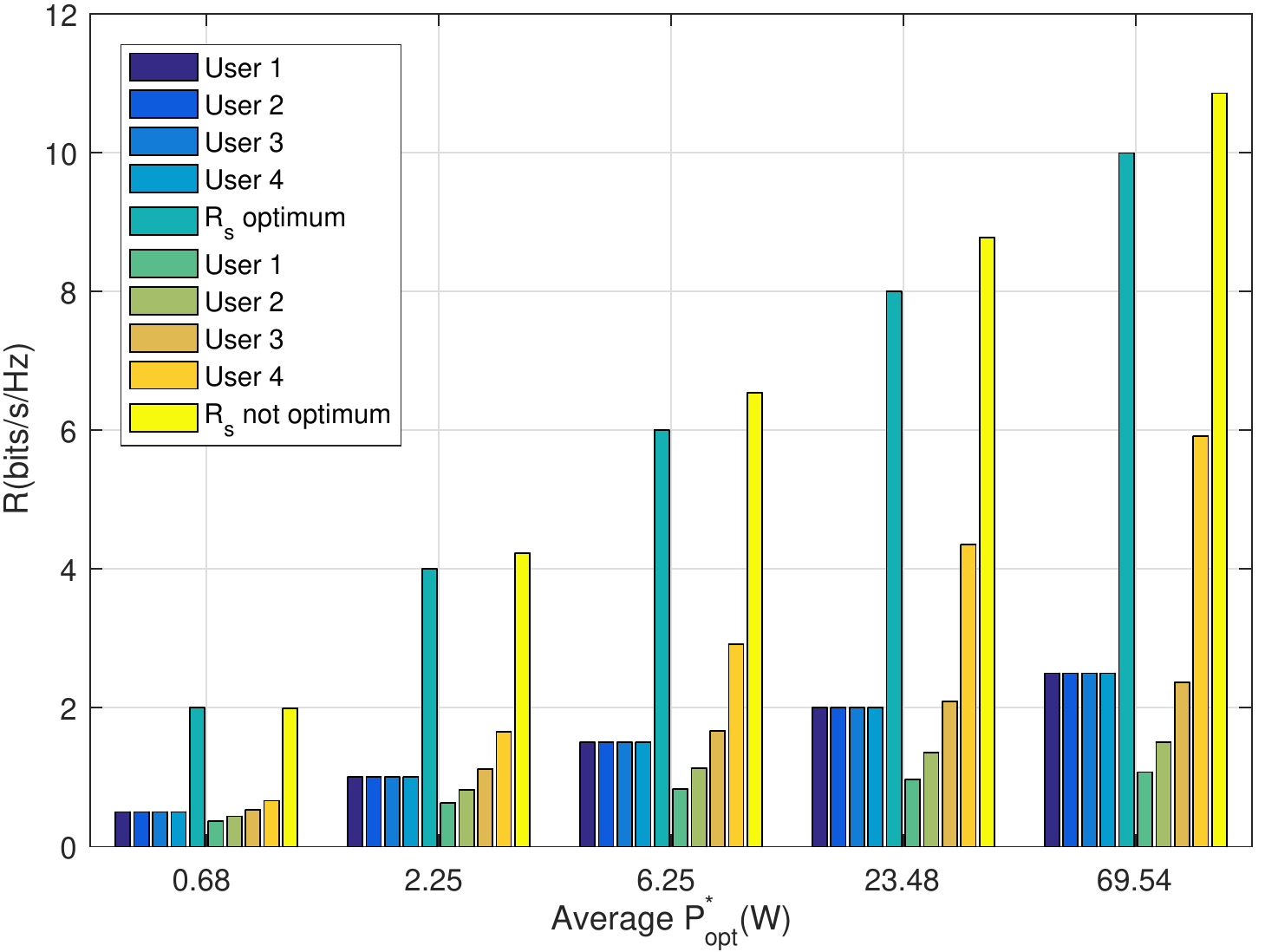}\\
	(a) Scenario 1  \hspace{66mm}  (b) Scenario 2 
	\vspace{-3mm}
	\caption{Data rates {\it versus} allocated power considered LoS pathloss ($\alpha=2$), problem \eqref{ConstraintsOptimizationBETA}. The non-optimal rates were obtained with the power proportion given by eq. (\ref{channelInverse}).} 
	\label{fig:RxPscenario1}
\end{figure}

To demonstrate the effectiveness of the proposed ERPA optimization following the rate fairness principle (same rate for all users), Fig. \ref{fig:RxPscenario1} depicts the respective optimum and non-optimum sum-rate.
 It was considered four users with  fixed rates equal to $0.5$, $1$, $1.5$, $2$ e $2.5$ bits/s/Hz and optimum average powers for Scenario $1$ with $\alpha=2$ (LoS - line of sight) for each user. The distributed {\it equal-rate optimal power allocation} (ERPA) strategy is compared with the one obtained by eq. (\ref{channelInverse}) with the same total power. It was observed that only the optimal system kept the same rates for all users at the cost of marginally reduced sum-rate at each total power scenario. In the non-optimal system, the user $4$ is closer to the BS and receives a greater proportion of power. This generates greater interference in other users, reducing their rates and increasing its own rate, which results in a bit higher sum rates. Notice that to ensure fairness, the strongest user (nearest) partially loses rate, while far away users gain rate by allocating more power for them; this mechanism happens for all the power-rate scenarios analyzed. This is the price paid to have fairness among users. Indeed, to maintain a rate of $R=2$ bits/s/Hz for all users in Scenario $1$ and $\alpha=2$, the total power required is $8.03$ W, Fig. \ref{fig:RxPscenario1} ; for the same case in Scenario $2$ the total power required is $23.48$ W. In Scenario $2$, the total power has to be $2.9$ times greater because users are more distant.
\begin{figure}[!htb]
	\centering
	\includegraphics[width=0.7\textwidth]{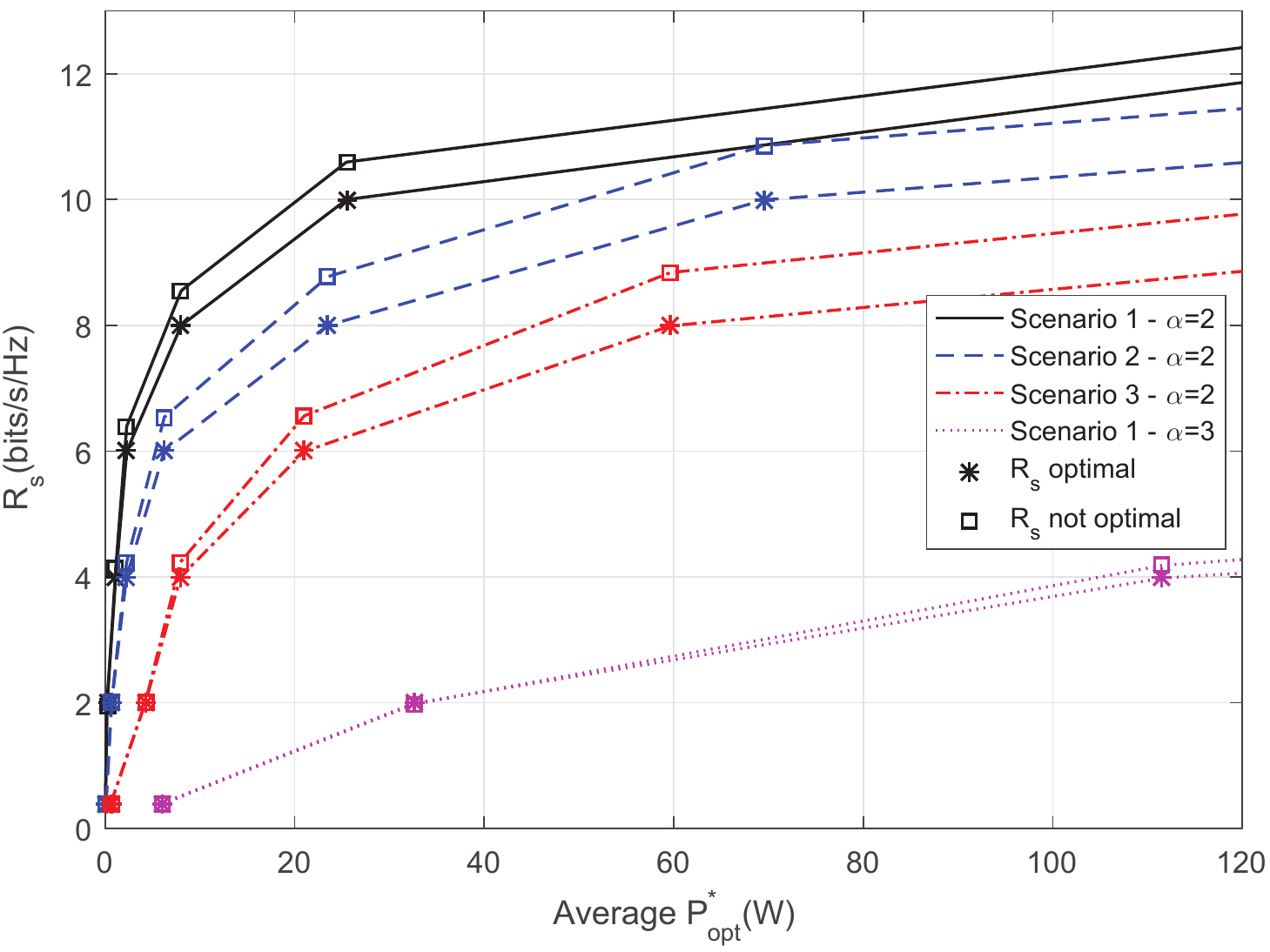}
	\vspace{-3mm}
	\caption{$R_s$ {\it versus} $P$ to $\alpha=2$ and $3$. The non-optimal rates were obtained with the power proportion given by equation (\ref{channelInverse}), problem \eqref{ConstraintsOptimizationBETA}.
		\label{fig:RsxPa2a3}}
\end{figure}

Fig. \ref{fig:RsxPa2a3} shows the sum-rate for optimal and a non-optimal distribution of total powers, for scenarios with different distances between users, and for $\alpha=2$ and $3$. With the same total power, the sums of non-optimal rates are somewhat larger for the non-ideal energy distribution because the closest users, with distances $d_2$, $d_3$, and $d_4$ have a higher power proportion and consequently higher rate than the farthest user, with distance $d_1$. For instance, in the environment with $\alpha=2$ and Scenario $3$, the sum of non-optimal rates with an average power budget of  $P_{opt}^*\approx 59.7$ W is $R_s=8.8$ bits/s/Hz while for the sum of optimal rates is $R_s=8.0$ bits/s/Hz for the same power budget of $59.7$ W.

\begin{figure}[!htb]
\centering
\includegraphics[width=0.49\textwidth]{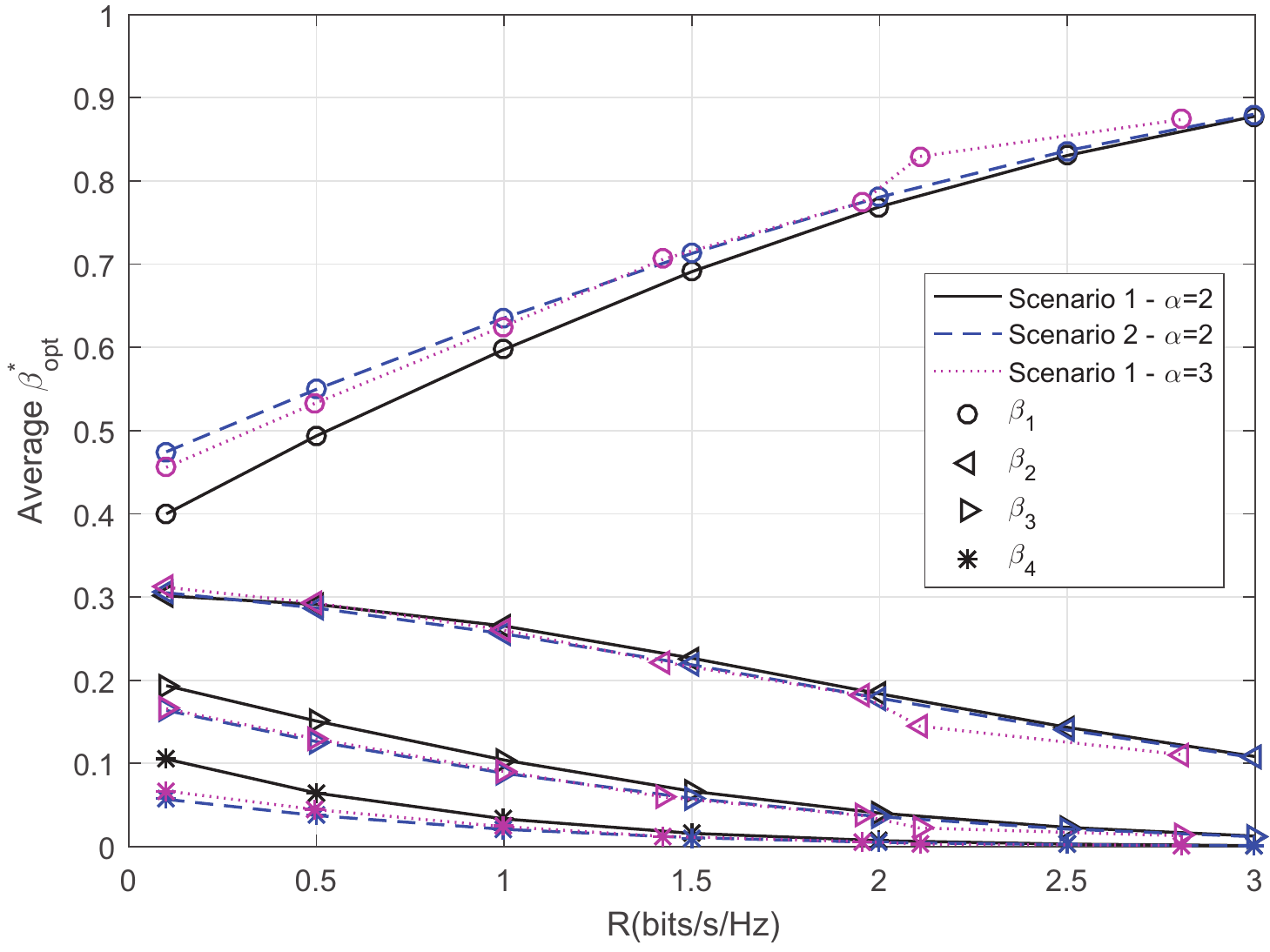}
\includegraphics[width=0.49\textwidth]{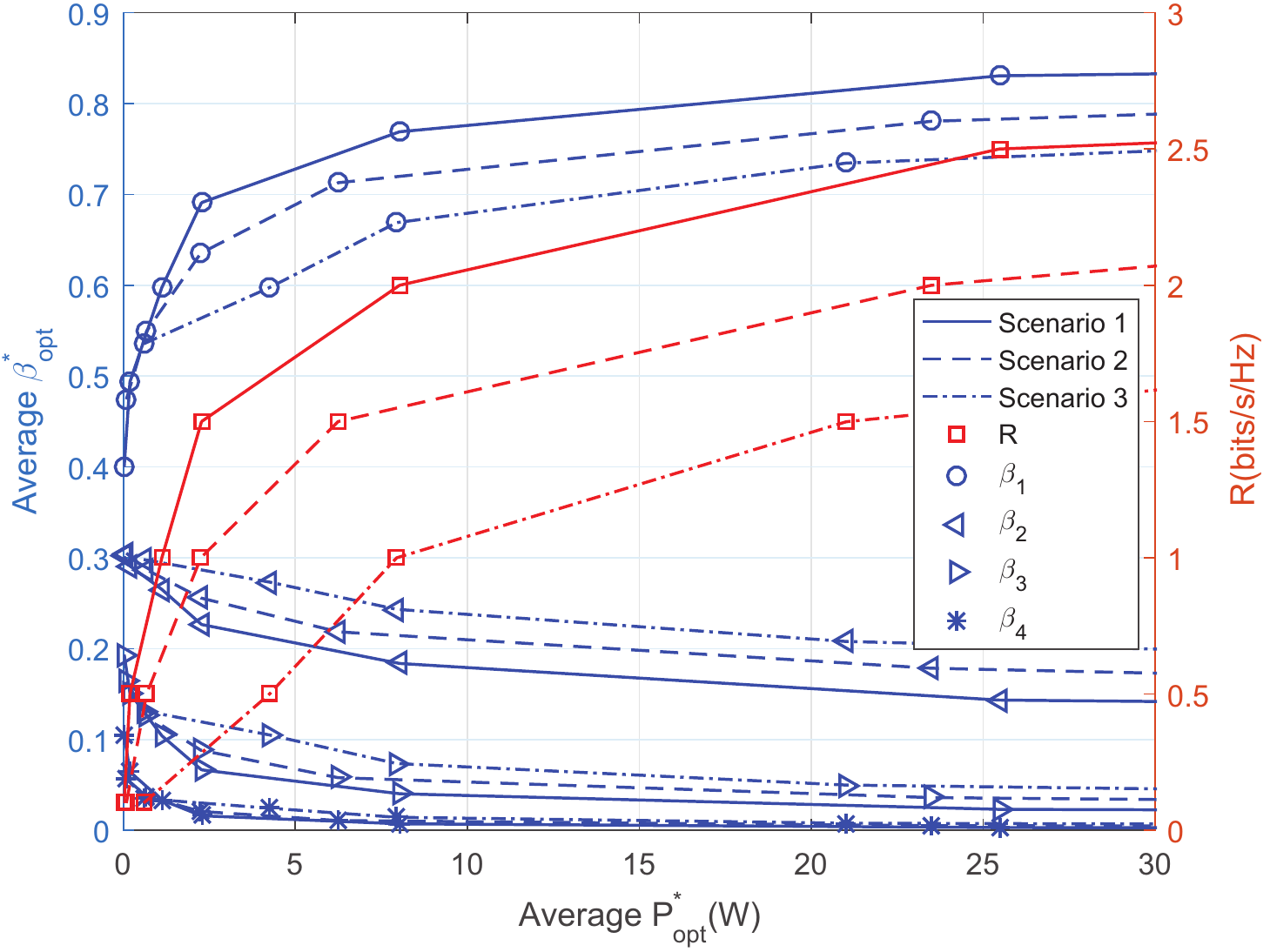}
\\
(a) $\alpha=2$ and $3$. \hspace{66mm} (b) $\alpha=2$.
\vspace{-3mm}
\caption{Optimal power proportion and sum-rate behavior, problem \eqref{ConstraintsOptimizationBETA}, for the three scenarios and different path loss coefficient: a) $\beta_{opt}^*\times R$; b) $\beta_{opt}^*\times P_{opt}^*$ and $R\times P_{opt}^*$.}
\label{fig:BETAxRa2a3}
\end{figure}

In Figure \ref{fig:BETAxRa2a3}(a), we have the variation of the power ratios of users $\beta_{opt}^*$ as a function of the rate of each user given by $R$ in bits/s/Hz. The power proportions for users of Scenario $2$ with path loss coefficient $\alpha=2$ and Scenario $1$ with $\alpha=3$ have a marginal difference, because the increase of distances of users in the first one is compensated by the increased path loss coefficient of the second one. For users of Scenario $1$ and the path loss coefficient $\alpha=2$, the power ratio of users is smaller because of decrease in the path loss effect.

Figure \ref{fig:BETAxRa2a3}(b) demonstrates how the power proportions $\beta_i$ behaves concerning the total power $P$ and the fixed-rate $R$ of each user. For a fixed rate of $R=1.5$ bits/s/Hz per user, the optimal total power is around $2.3$ W, $6.2$ W, and $21.0$ W for the Scenarios $1$, $2$ and $3$, respectively, while the power proportion for users $1$ are around $\beta_1=0.7,\; 0.71$ and $0.73$. The same power proportion, the different total power and the same data rate occur because of the distance differences among users for each of the three scenarios.

\subsection{Fairness}
Fairness index {$\mathcal{F}_{\textsc{it}}$} defined in \eqref{NewFairness} has been applied to the proposed ERPA-NOMA strategy for the four different scenarios. As shown in Table \ref{table:NewFairness} , fairness index results are very close to $1$ demonstrating the effectiveness of the proposed optimization method. The maximum value of $\mathcal{F}_{{\textsc{it}}}=0.976$ is in Scenario $3$ for $R=1$ bits/s/Hz and the minimum value of $\mathcal{F}_{{\textsc{it}}}=0.939$ is in Scenario $1$ for $R=3$ bits/s/Hz.
\begin{table}[!htb]
	\centering
	\caption{Fairness $\mathcal{F}_{{\textsc{it}}}$ obtained from the  equation (\ref{NewFairness}).}
	\begin{tabular}{|c|c|c|c|c|}
		\hline
		\multicolumn{1}{|l|}{} & \multicolumn{3}{c|}{$\alpha=2$}                                   & $\alpha=3 $           \\ \hline
		\textbf{R(bits/s/Hz)}  & \textbf{Scenario 1} & \textbf{Scenario 2} & \textbf{Scenario 3} & \textbf{Scenario 1} \\ \hline
 		\textbf{1.0}           & 0.947               & 0.968        &\underline{{0.976}} & 0.953               \\ \hline
		\textbf{1.5}           & 0.943               & 0.962               & 0.970               & 0.948               \\ \hline
		\textbf{2.0}           & 0.941               & 0.958               & 0.966               & 0.945               \\ \hline
		\textbf{2.5}           & 0.940               & 0.955               & 0.962               & 0.943               \\ \hline
		\textbf{3.0}   &\underline{0.939} & 0.952               & 0.960               & 0.942               \\ \hline
	\end{tabular}
	\label{table:NewFairness}
\end{table}

 Comparing the Scenarios $1$ through $3$ for $\alpha=2$, one can conclude that the increase in fairness index is gradual, approaching to 1 when the sum of users' distance increases (Scenario 3). Moreover, the fairness index $\mathcal{F}_{{\textsc{it}}}$ increases when the non-line of sight occurs, i.e. when $\alpha> 2$ happens. Finally, $\mathcal{F}_{{\textsc{it}}}$ decreases slowly when the equal rate per user $R$ increases, and it demonstrates how difficult it is to attain more restrictive tied QoS goals.

Jain's fairness index $\mathcal{F}_j$ \eqref{fairness} resulted in $1$ once it was ensured that all users had the same rate in optimization process. This was sufficient to ensure maximum fairness when calculated for the index. However, fairness index $\mathcal{F}_{{\textsc{it}}}$ based on the information theory \cite{Gui2019} resulted close to 1. For index $\mathcal{F}_{{\textsc{it}}}$ to be equal to 1, powers must also be uniformly distributed.
%

\subsection{Energy Efficiency Optimal Operation Points}
\begin{figure}[!htbp]
	\centering
	\includegraphics[width=0.49\textwidth]{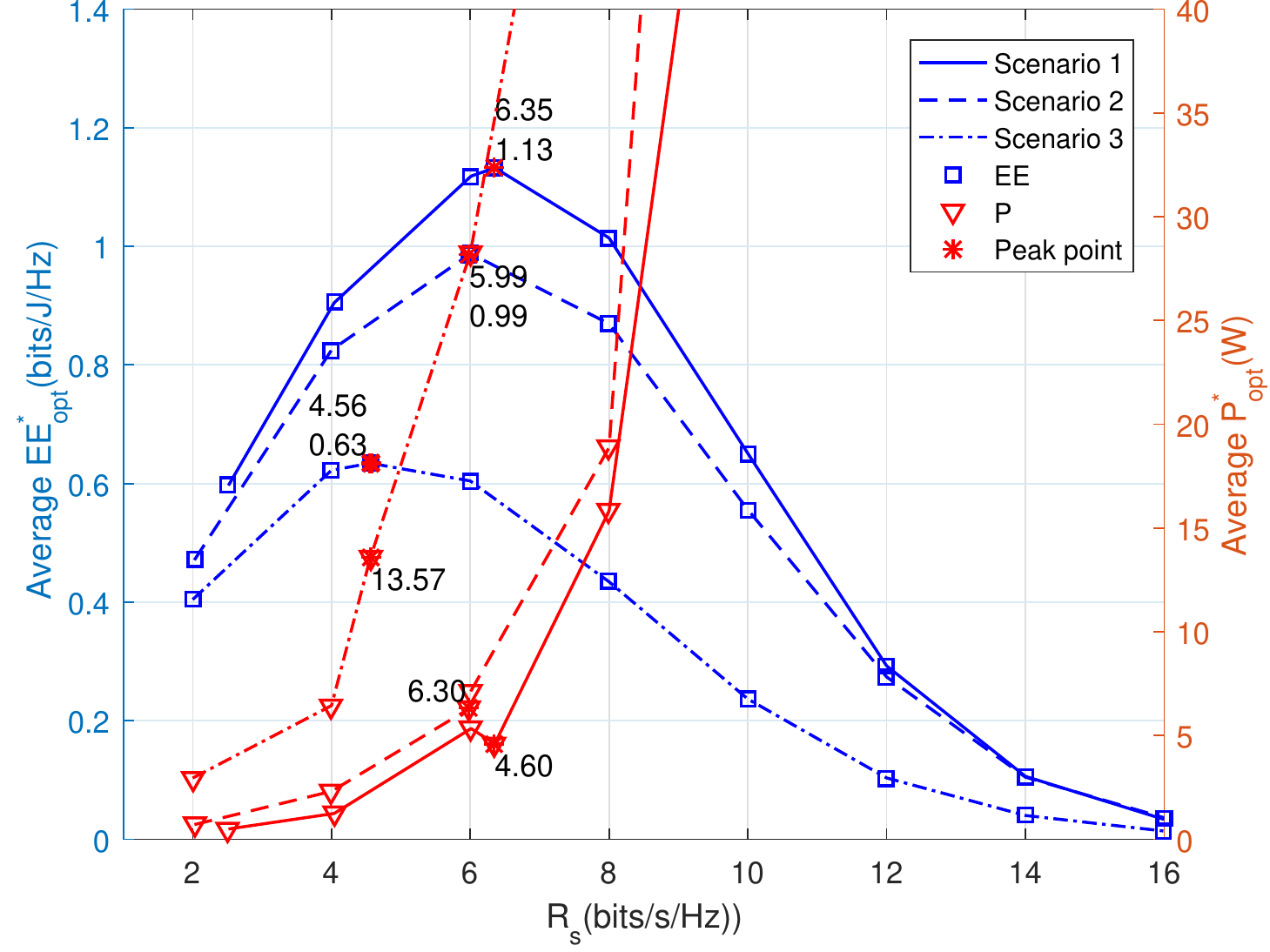}
	\includegraphics[width=0.49\textwidth]{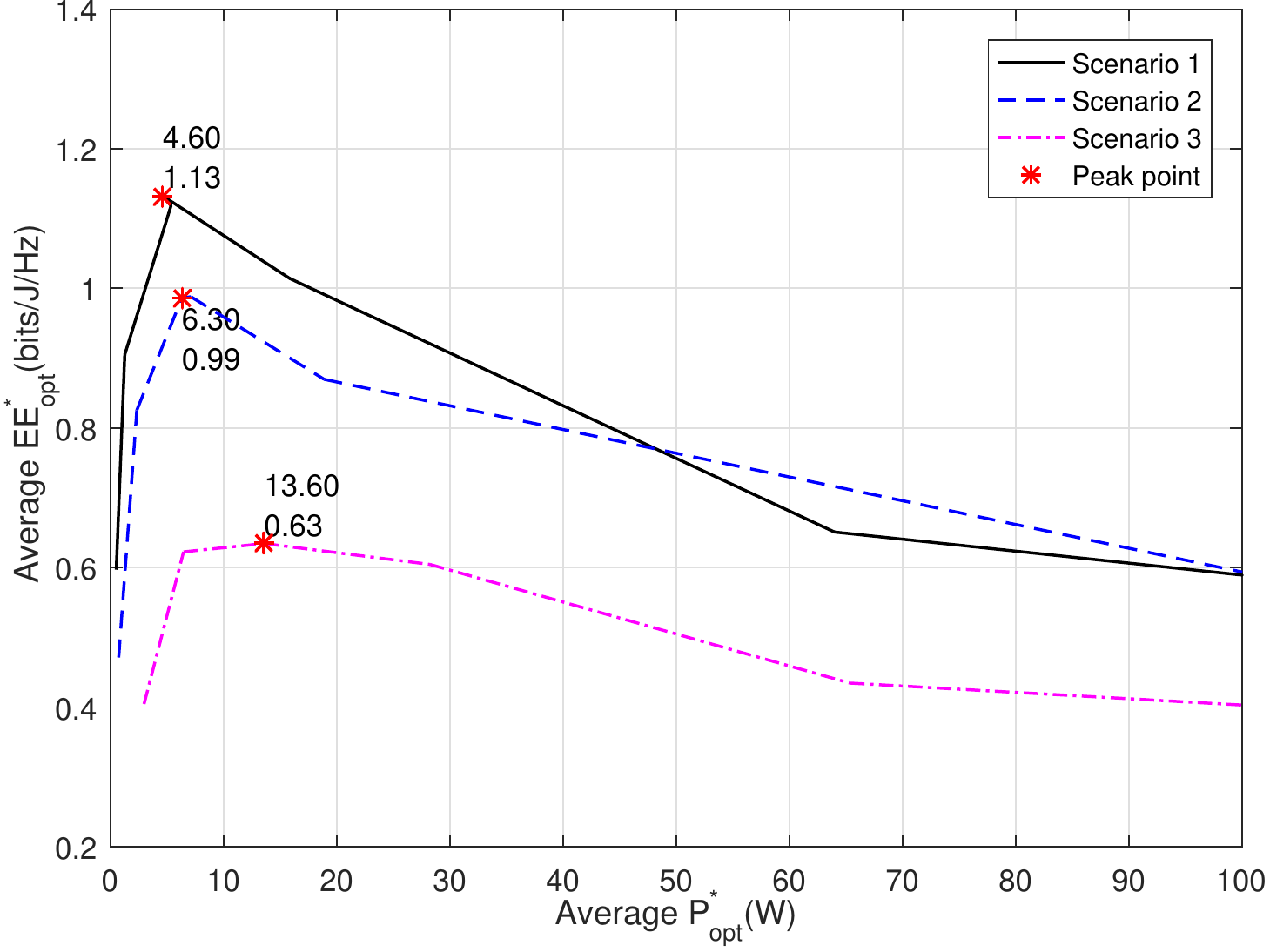}
	(a) $\text{EE}_{opt}\times R_s$ and $P_{opt}\times R_s$ \eqref{ConstraintsOptimizationBETApEE}.  \hspace{35mm} (b) $\text{EE}_{opt}\times P_{opt}$ \eqref{ConstraintsOptimizationBETApEE}.
	\vspace{-3mm}
	\caption{Optimal points considering $\alpha=2$. Peak point problem \eqref{ConstraintsOptimizationBETApReEE}.}
	\label{fig:EExRsumPa2}
\end{figure}

The optimal point for energy efficiency (EE$^*$) is depicted in Figure \ref{fig:EExRsumPa2}(a) and (b) for the three fixed-value scenarios and for line of sight channel $\alpha=2$. For Scenario $1$, EE$^*=1.13$ bits/ J/ Hz operation point results in a rate of $R=6.35$ bits/s/Hz for a total power consumption of $P=4.60$ W. While for Scenarios $2$ and $3$ we have rates of $R=5.99$ bits/s/Hz and $R=4.56$ bits/s/Hz for a total power of $P=6.30$ W and $P=13.57$ W, which resulted in $\text{EE}^*=0.99$ bits/J/Hz and $\text{EE}^*=0.63$ bits/J/Hz, respectively. Of course, Scenario $1$ (closer users), less power is required to reach the specified rate, so that very low power reaches the highest rate and consequently higher EE$^*$ as it can be seen in graph of Figure \ref{fig:EExRsumPa2}(a) and (b) in the points in asterisk. Indeed, the EE$^*$ point for Scenario $1$ is higher than users of Scenarios $2$ and $3$, since users of the Scenario $1$ are closer each other, which can be corroborated by the behavior of  Figure \ref{fig:EExRsumPa2}(b).

\subsection{Resource Efficiency  (EE $\times$ SE) Optimal Point as a function of the Number of Users}
\begin{figure}[!htbp]
\centering
\includegraphics[width=0.49\textwidth]{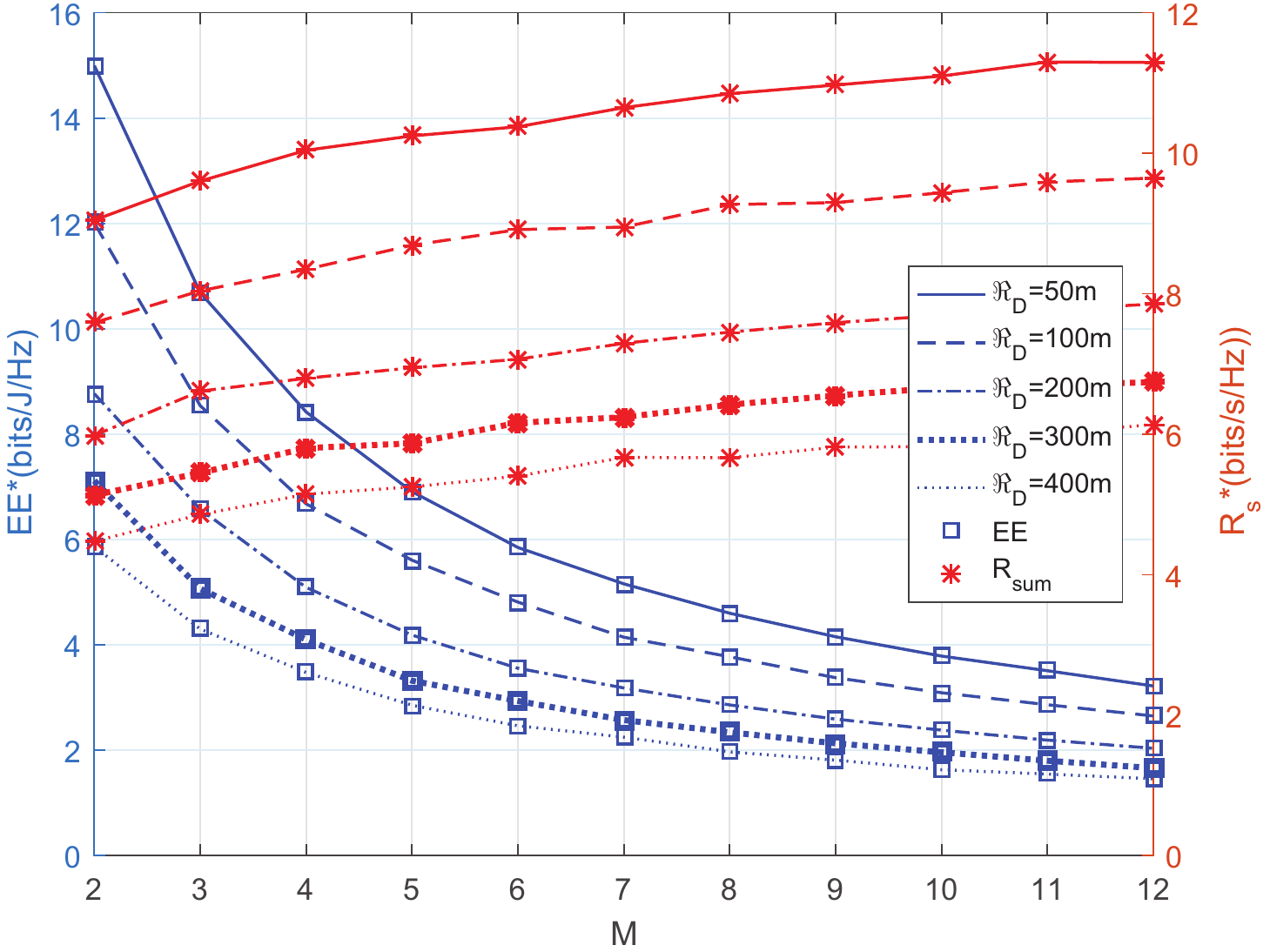}
\includegraphics[width=0.49\textwidth]{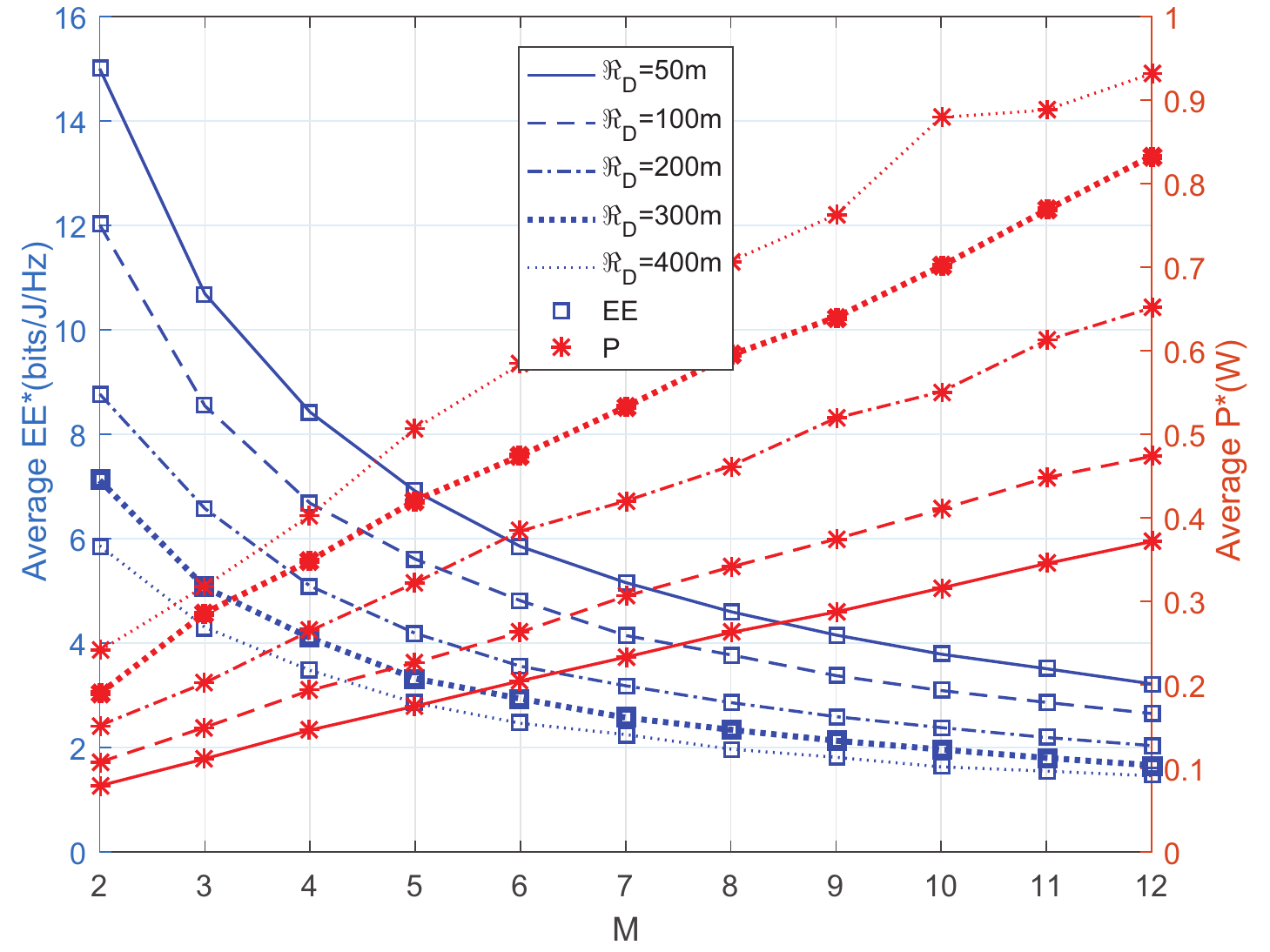}
	\\
(a) $\text{EE}_{opt}^*\times M$ and $R_s\times M$ \hspace{35mm} (b) $\text{EE}_{opt}^*\times M$ and $P_{opt}^*\times M$	
\vspace{-3mm}
\caption{Trade-off between EE and SE for some radius $\Re_D$ and $\alpha=2$, problem \eqref{ConstraintsOptimizationBETApReEE}.}
\label{fig:EEsumRATEusera2h}
\end{figure}

\begin{figure}[t]
\centering
\includegraphics[width=0.49\textwidth]{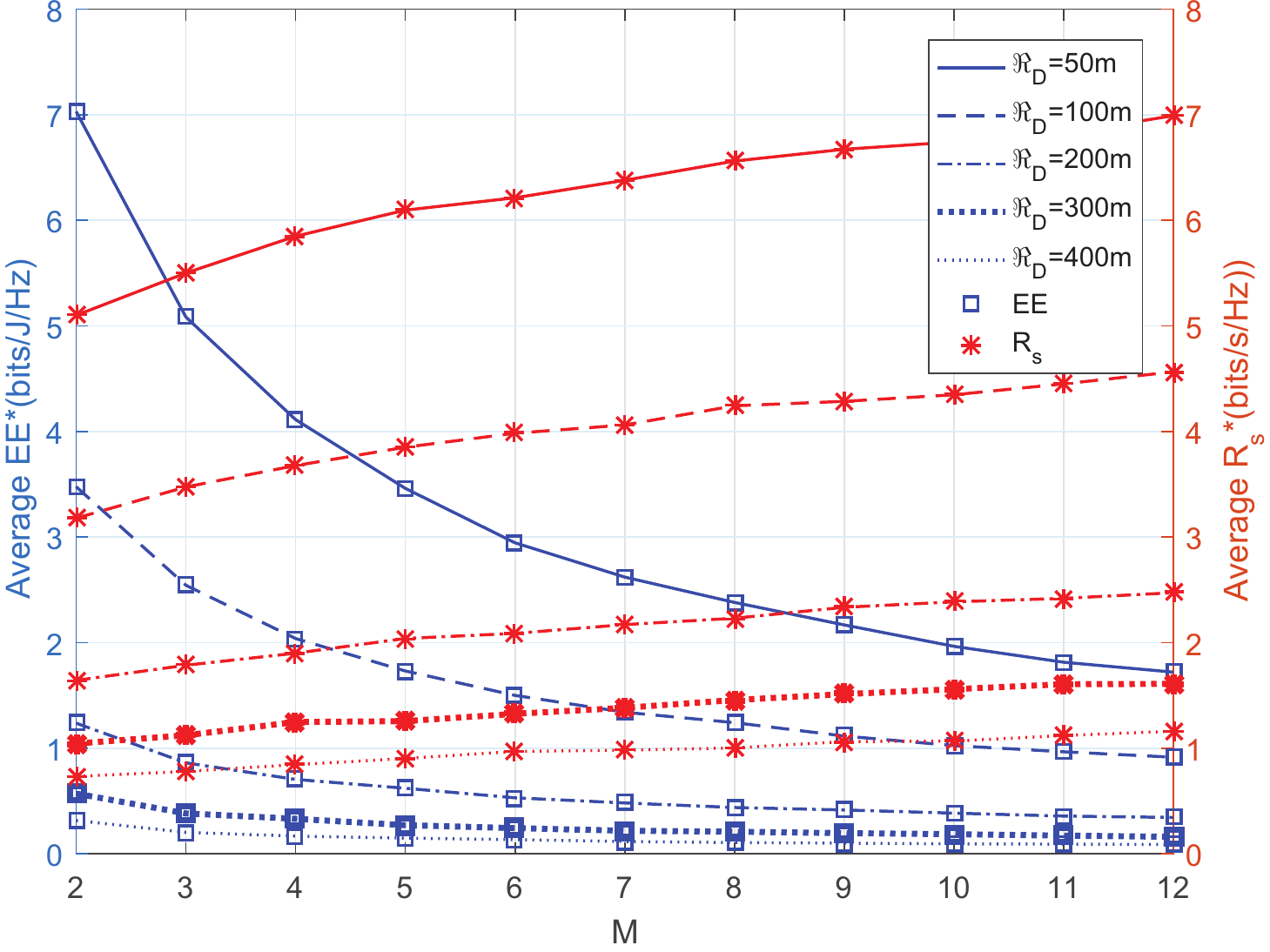}
\includegraphics[width=0.49\textwidth]{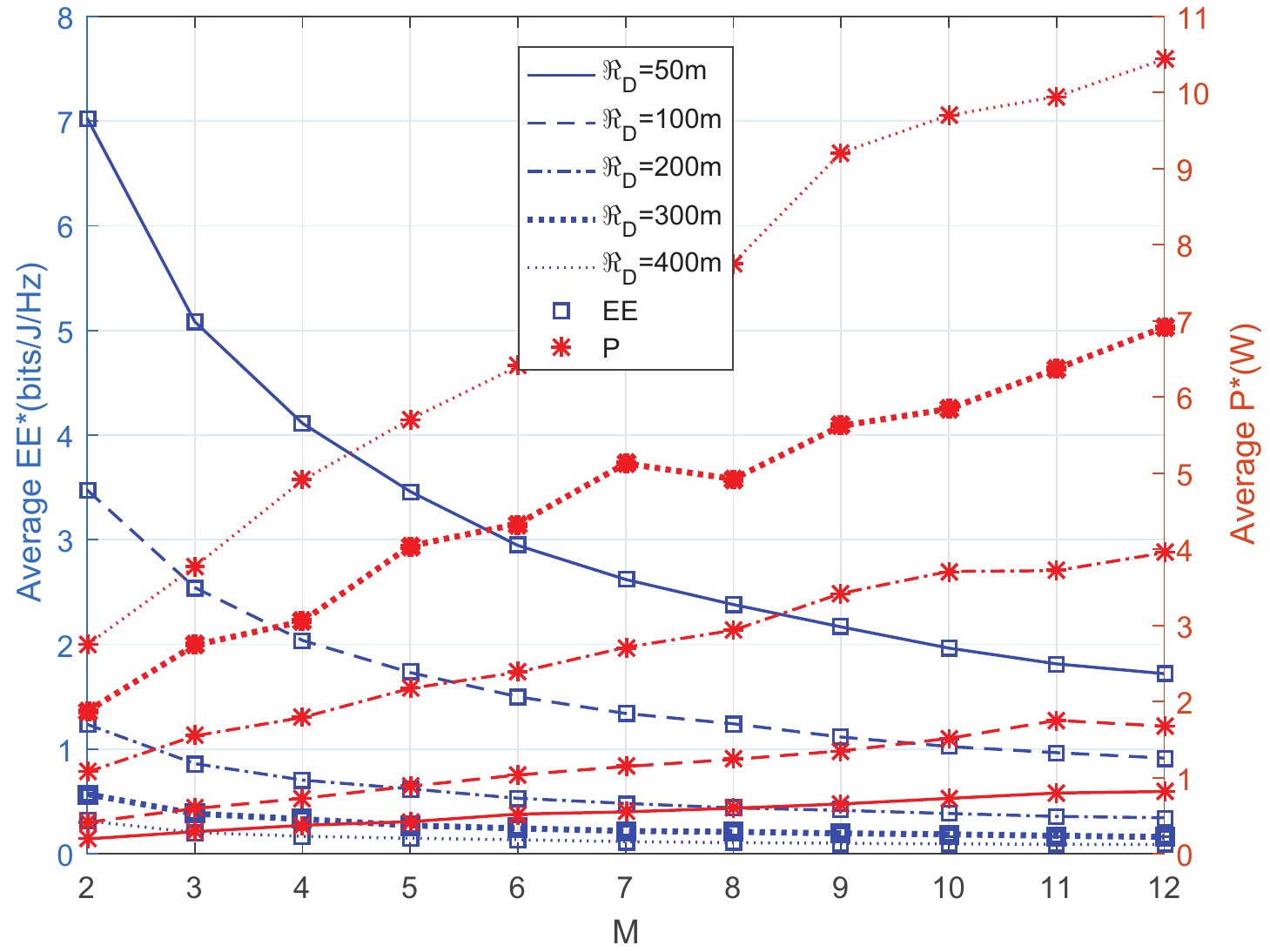}
	\\
(a) $\text{EE}_{opt}^*\times M$ and $R_s\times M$ \hspace{46mm} (b) $\text{EE}_{opt}^*\times M$ and $P_{opt}^*\times M$	
\vspace{-3mm}
\caption{Trade-off between EE and SE for some radius $\Re_D$ and $\alpha=3$, problem \eqref{ConstraintsOptimizationBETApReEE}.}
\label{fig:EEsumRATEusera3h}
\end{figure}

\begin{figure}[t]
	\centering
	\includegraphics[width=0.49\textwidth]{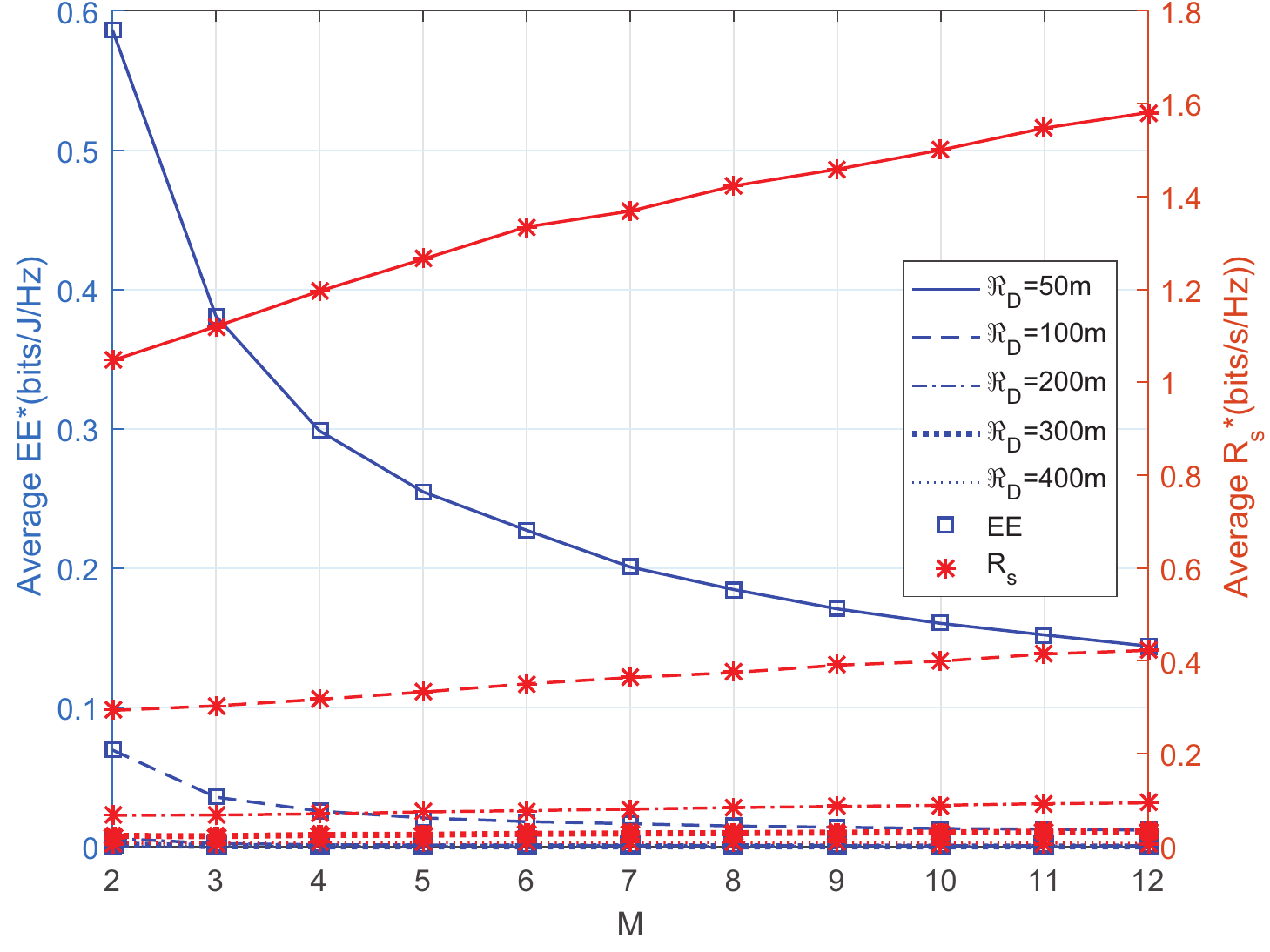}
	\includegraphics[width=0.49\textwidth]{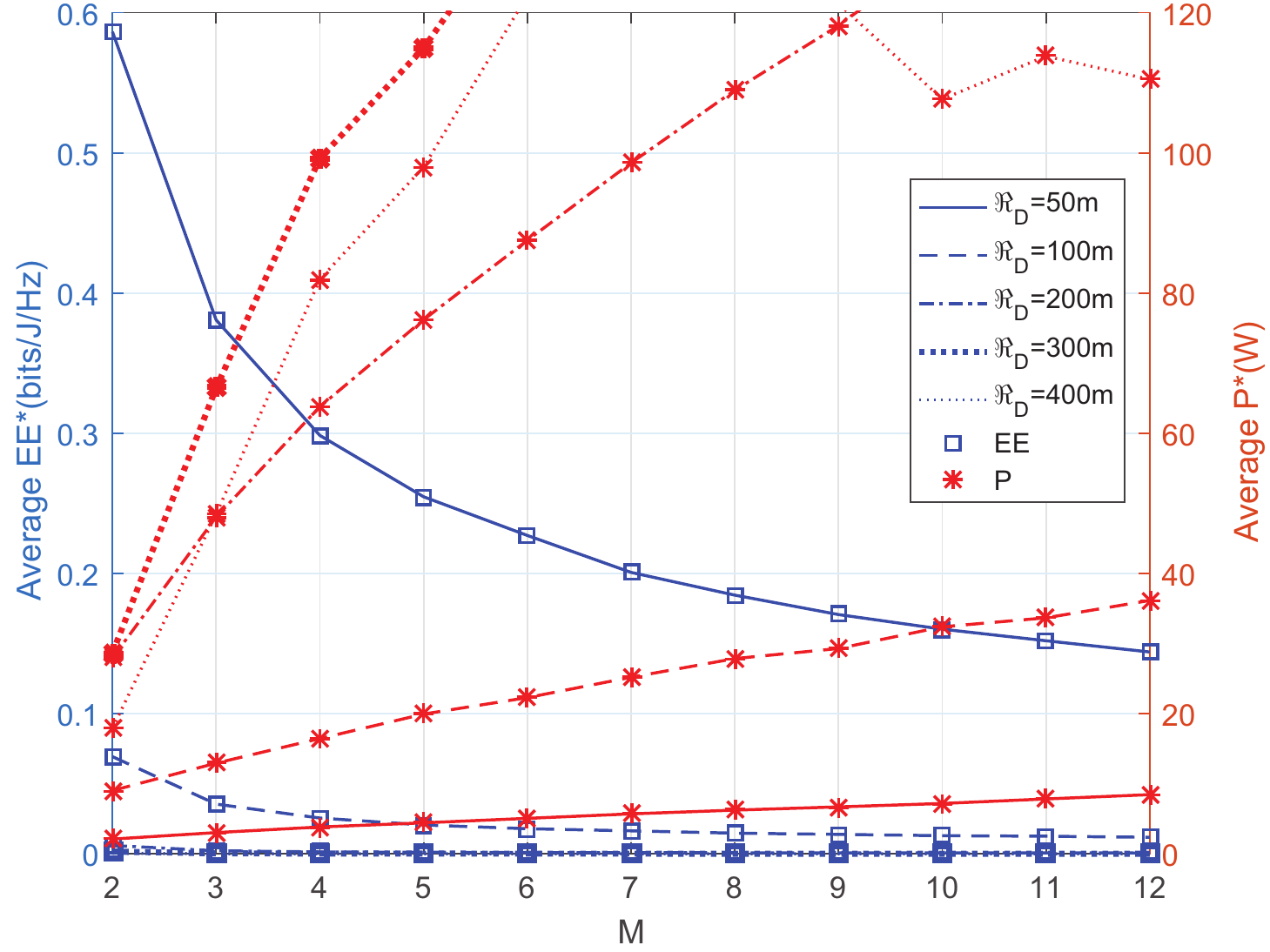}
		\\
	(a) $\text{EE}_{opt}^*\times M$ and $R_s\times M$ \hspace{46mm} (b) $\text{EE}_{opt}^*\times M$ and $P_{opt}^*\times M$	
	\vspace{-3mm}
	\caption{Trade-off between EE and SE for some radius $\Re_D$ and $\alpha=4.5$, problem \eqref{ConstraintsOptimizationBETApReEE}.}
	\label{fig:EEsumRATEusera45h}
\end{figure}

Problem \eqref{ConstraintsOptimizationBETApReEE} was used to obtain RE from EE. As it can be seen from the graph in Figure \ref{fig:EExRsumPa2} , the peak point is very close to maximum point to fixed $R$ in problem \eqref{ConstraintsOptimizationBETApEE}. In Figures \ref{fig:EEsumRATEusera2h} , \ref{fig:EEsumRATEusera3h} and \ref{fig:EEsumRATEusera45h} , one can see the optimal values for energy efficiency, sum rates and total power as a function of the number of users $M$ from the perspective of the trade off EE $\times$ SE. In this case, the sum rate $R_s$ and total power $P$ are considered as variables in the optimization process of problem \eqref{ConstraintsOptimizationBETApReEE}. To obtain the curves, it was considered that users have uniform distribution within the cell of radius $\Re_D$ and $P_c=250m$W per user\cite{Devroye1986}. We considered $10^4$ samples for each number of users $M$. Table \ref{table:TradeoffEExRsa2} summarized the relevant points of numerical results obtained from Figure \ref{fig:EEsumRATEusera2h} , \ref{fig:EEsumRATEusera3h} and \ref{fig:EEsumRATEusera45h} , where the best trade off between EE and $R_s$ were selected for each scenario. This trade off represented the number of users that obtained the best system resource efficiency (RE) to $\xi_0=1.8$ in Eq. \eqref{RE}. 

For different rays $\Re_D$ and $\alpha$, two users/cluster configuration showed the best RE as seen in Table \ref{table:TradeoffEExRsa2} . EE and SE reduced with the increase of rays of cells and consequently the increase of optimum powers obtained due to larger distances of users. These values were the best configuration to obtain the highest resource efficiency in NOMA system in the presented scenarios.


	\begin{table}[!h]
	\begin{center}
		\caption{Trade-off point between EE and $R_s$ for some $\Re_D$ and users with uniform distributions.}
		\begin{tabular}{|c|c|c|c|c|c|c|c|c|c|c|c|c|}
			\hline
			\multicolumn{1}{|l|}{} & \multicolumn{4}{c|}{$\alpha=2$}                                  & \multicolumn{4}{c|}{$\alpha=3$}                                  & \multicolumn{4}{c|}{$\alpha=4.5$}                                \\ \hline
			\textbf{$\Re_D$}       & \textbf{User} & \textbf{EE} & \textbf{$R_s$} & \textbf{$P$} & \textbf{User} & \textbf{EE} & \textbf{$R_s$} & \textbf{$P$} & \textbf{User} & \textbf{EE} & \textbf{$R_s$} & \textbf{$P$} \\ \hline
			50                     & 2             & 15.00       & 9.05           & 0.08              & 2             & 7.03        & 5.10           & 0.20              & 2             & 0.59        & 1.05           & 2.11              \\ \hline
			100                    & 2             & 12.01        & 7.59           & 0.11              & 2             & 3.48        & 3.18           & 0.41              & 2             & 0.07        & 0.29           & 9.03              \\ \hline
			200                    & 2             & 8.77        & 5.98           & 0.15              & 2             & 1.24        & 1.64           & 1.07              & 2             & 0.006       & 0.07           & 27.94             \\ \hline
			300                    & 2             & 7.12        & 5.13           & 0.19              & 2             & 1.04        & 0.57           & 1.87              & 2             & 0.0005      & 0.02           & 28.65            \\ \hline
			400                    & 2             & 4.48        & 5.86           & 0.24              & 2             & 0.32        & 0.73           & 2.74              & 2             & 0.0005      & 0.008          & 17.92            \\ \hline
		\end{tabular}
		\label{table:TradeoffEExRsa2}
		\end{center}
	\end{table}


\newpage
\section{Conclusions}
In this work, we defined the proportion of energy distributed optimally for each user to a given transmission rate in a NOMA system of two users or more. In our design, it was also possible to optimize the total power for the same rate {maximum fairness among} all users. We used the techniques of SQP {which presents fast convergence feature. The numerical results corroborate the minimum total power allocation for two or more users need} to reach the defined capacity. The {design and analysis were developed aiming to} find the minimum equal rate for each user in each scenario, with the maximum energy efficiency. Maximum fairness was guaranteed in our project. We also found the trade-off point between EE and the sum-rate, that is, the system operating in the point of best resource efficiency.

\end{document}